\documentclass[11pt]{article}

\usepackage{graphicx,amsfonts,amsmath,overcite,amsmath}

\begin{document}

\title{Mathematical Models of Bipolar Disorder}

\author{\\ Darryl Daugherty \\ Department of Mathematics \& Statistics \\ 
California State Polytechnic University,
Pomona \\ \\ Tairi Roque-Urrea \\ Department of Mathematical Sciences \\
Binghamton University \\ \\ John Urrea-Roque \\ Department of
Mathematical Sciences \\ Binghamton University \\ \\ Jessica
Snyder \\ School of
Mathematics \\ Georgia Institute of Technology \\ \\ Stephen Wirkus \\
Department of Mathematics \& Statistics \\ California State
Polytechnic University, Pomona\\ \\ Mason A. Porter \\ School of
Mathematics \& Center for Nonlinear Science \\ Georgia Institute
of Technology \\ \\ }

\maketitle



\section*{Abstract}

We use limit cycle oscillators to model Bipolar II disorder, which is 
characterized by alternating hypomanic and
depressive episodes and afflicts about one percent of the United
States adult population. We consider two nonlinear oscillator
models of a single bipolar patient. In both frameworks, we begin
with an untreated individual and examine the mathematical effects
and resulting biological consequences of treatment. We also
briefly consider the dynamics of interacting bipolar II
individuals using weakly-coupled, weakly-damped harmonic
oscillators. We discuss how the proposed models can be used as a
framework for refined models that incorporate additional
biological data.  We conclude with a discussion of possible
generalizations of our work, as there are several
biologically-motivated extensions that can be readily incorporated
into the series of models presented here.


\section{Characterizations of Bipolar II Disorder}

\indent

About one percent of the United States adult population is
afflicted with bipolar disorder (manic depression),\cite{Management1} which 
poses myriad difficulties to clinical practitioners.  It is difficult to
diagnose, as bipolar patients often do not adhere to treatment
and/or medication, and most drugs can be toxic if taken
individually.\cite{treat}  The available data on bipolar disorder
is sparse---due in part to a lack of agreement on well-suited
trial design---so it is very difficult to study using clinical
trials.\cite{post}

Psychiatrists have established a broad range of criteria for
classifying bipolar disorder; see, for example, the Diagnostic and
Statistical Manual of Mental Disorders, Fourth Edition (DSM
IV).\cite{Diagnostic}  Its characteristics, which need not all be
present in every bipolar individual, include {\it mixed episodes},
in which it is possible to simultaneously experience symptoms of
both mania and depression, and {\it rapid cycling}, in which a
patient experiences at least four cycles per
year.\cite{Management1}

There are two primary types of bipolar disorder. {\it Bipolar I
disorder} is characterized by a combination of manic and
depressive episodes with the possibility of mixed episodes,
whereas {\it bipolar II disorder} is characterized by a
combination of hypomania\footnote{The main distinction between hypomanic 
episodes and manic ones is that the former are much less severe than the 
latter.  Additionally, mania can last much longer than a week, whereas 
hypomania has been shown to have a median duration of about 
one-two days\cite{iain2} and need only last four days to reach DSM-IV 
criteria.\cite{Diagnostic}} and depressive episodes.\cite{Diagnostic,bip}  
Bipolar II patients tend to be more prone to rapid cycling---especially if 
initially treated only with antidepressants.\cite{treat,Wehr1}  In this paper, 
we focus on bipolar II disorder for two reasons.  First, individuals who 
suffer from it are more often observed to exhibit approximately periodic 
mood swings than bipolar I individuals.  Second, bipolar II individuals do 
not experience mixed episodes.

Bipolar II disorder is often misdiagnosed as either unipolar
depression or a severe personality disorder.\cite{iain} To be
correctly diagnosed, a patient seeking treatment must give an
accurate description of his past behavior. Because of the nature
of hypomanic episodes (increased energy, decreased need for sleep,
{\it etc.}) patients often do not describe these conditions to
doctors and are therefore diagnosed with unipolar
depression.\cite{fawcett} According to the Results of the National
Depressive and Manic-Depressive Association 2000 survey of people
with bipolar disorder, over one third of respondents sought
professional help within a year of onset of symptoms. However, it
took as many as ten years and four physicians for some patients to
be correctly diagnosed.\cite{vornik}

Treatment for bipolar disorder ideally includes a combination of
medication and therapy. Typical drug treatments include mood
stabilizers, antipsychotics, antidepressants, and select
anticonvulsants.  Among the more commonly used drugs are Lithium,
Valproate (also know as Depakote), Carbamazepine (also known as
Tegretol), and Prozac. Mood controlling drugs such as Lithium take
four to ten days to reach therapeutic levels in the blood stream,
so initial treatment is likely to include both antidepressants and
antipsychotics.\cite{bip} During the maintenance state,
antidepressants and antipsychotics are generally supplemented by
mood stabilizers. Monotherapy (i.e., single-drug therapy) is
generally avoided by clinicians because of the strong side effects
of some of the drugs used for treatment.  Other drugs that have
been employed include selective serotonin reuptake inhibitors
(SSRI) and monoamineoxidase inhibitors (MAOI), both of which are
typically used for depression.  In some cases, special
care must be taken to ensure that a bipolar individual does not
fall into a pattern of rapid cycling or become addicted to the
medication used for treatment.  In fact, substance abuse has been 
associated with bipolar disorder.\cite{bip}

Bipolar II disorder is also known to be highly heritable. It has
been reported, for example, that the offspring of people with
bipolar II disorder have a $35$ percent chance of being afflicted
as well.  In particular, there are known cases of family units
with multiple bipolar II individuals, so the dynamics of
interacting bipolar patients is also of interest to
psychologists.\cite{Judd,weiss,colom}

In this paper, we propose and examine two mathematical models that
attempt to represent the qualitative dynamics of individual
bipolar patients.  We also consider a closely interacting pair of
bipolar patients (which can occur, for example, in households with
bipolar sibling pairs\cite{sibling}).

The importance of the present work is that it suggests ways of
thinking about bipolar disorder.  In particular, mathematical models 
have the potential to provide significant
insight into this disorder, provided there is adequate data to
constrain the models.  Studies using time-series analysis (of a
relatively small number of bipolar and normal individuals) suggest
that relatively simple mechanisms may be responsible for the
complex mood variations in bipolar disorder.\cite{gott}  The
models we develop provide a first step toward using mathematical
modeling to increase scientific understanding of this disorder.
Ultimately, our models will need refinement to allow greater
predictability and ties to candidate biological processes.  Our
hope is for the present work to motivate the collection of the additional
time-series data necessary to achieve this goal.

\section{Scope of the Paper}

\indent

Numerous research articles discuss bipolar disorder, focusing
primarily on clinical trials and case studies.  Unfortunately,
very little of this prior research is amenable to mathematical
treatment, whose employment could prove vital in advancing our
knowledge of this medical condition.\cite{iain}  Nevertheless,
several studies of bipolar disorder---including a recent
comprehensive article by Judd, {\it et al.} (2003)---characterize
the moods of bipolar patients in terms of the fraction of weeks
during a year when particular symptoms were (or were not)
displayed.\cite{Judd}  While the models we examine in the present
paper can also be interpreted in these terms, a weekly or daily
time-series description of the symptoms could inspire the creation
or modification of mathematical models that may shed additional
light on the underlying dynamics of bipolar disorder. For example,
early work by Wehr and Goodwin (which demonstrates oscillatory
behavior in the moods of bipolar individuals) could be used, in
principle, to help refine our mathematical framework.  However,
their time-series description included only one bipolar patient,
so its practical utility in the development of a mathematical
description of bipolar II disorder is limited.\cite{Wehr1}

A comprehensive study (with a large number of clinical trials)
like that of Judd, {\it et al.} that uses the time-series format
of Wehr and Goodwin could lead to far more realistic mathematical
models than the available data currently allows.  Although
arduous, such an undertaking would yield important advancements in
modeling and understanding bipolar disorder.  Our goal in this
paper is considerably more modest---we use dynamical systems
theory to develop minimalist models of bipolar II disorder that we
hope will lead to additional qualitative understanding of the
behavioral oscillations associated with this medical condition.
The novelty of our work lies in our mathematical approach to the
modeling of bipolar disorder.

Although there exist biological models for this medical condition,
mathematical equations have never (to our knowledge) previously
been employed to study bipolar individuals. In this work, we
propose and analyze two mathematical models of bipolar II
disorder. Rather than focusing on the method of diagnosis (which is a 
difficult medical problem), we instead concentrate on the dynamics of 
our models under the proposed treatment strategy, which is understood to
involve a combination of drugs and/or therapy. The models we study
are not meant to represent a specific treatment but are instead
intended to provide some insight into the complicated dynamics of
bipolar disorder.  In this respect, we consider our work to be a
first step in developing mathematical models of the mood swings of
bipolar individuals.   The shift in thinking from a statistical
model to a nonlinear mathematical model for mood variations has
been suggested by some authors, including Totterdell {\it et
al.}\cite{Totterdell} Additionally, Ehlers\cite{ehlers} and
Gottschalk, {\it et al.}\cite{gott} have noted the importance of
nonlinear and chaotic dynamics and the employment of simple models
incorporating such features to achieve better understanding of
mood and behavior.  In fact, the latter authors attempted to
identify and quantify the nonlinear behavior in the mood of
bipolar patients by using time-series analysis to study power
spectra and fractal dimension.  However, they did not attempt to
develop mathematical equations to model mood swings in bipolar
individuals (although they did speculate that utilizing van der
Pol oscillators may be useful), which is the goal of this work.
In this respect, our theoretical work nicely complements the data
analysis of Gottschalk, {\it et al}.

Reinterpreting the time series results of Totterdell, {\it et al.}
and gathering similar time series data for bipolar patients has
considerable merit.  We hope that our work in developing dynamical
equations describing the mood swings of bipolar individuals, in
conjunction with  additional studies like those of Totterdell,
{\it et al.} and  Gottschalk, {\it et al.}, leads to the eventual
development of more detailed models that incorporate clinical data
from both bipolar and ``normal" individuals.

Numerous studies have proposed possible underlying mechanisms for
bipolar disorder, and some have even examined the effect of light
stimuli and seasonal variations.  Potentially insightful
mathematical models can be motivated from such work, but the perspective 
we take is to employ minimalist mathematical models describing biological and
physical oscillators that assume {\it a priori} the existence of
asymptotic oscillations (i.e., limit cycles).\cite{Lienard}  This
allows us to gain insight into the dynamics of the mood swings of
bipolar II individuals, despite the fact that the underlying
dynamics are not fully understood at the level of chemicals in the
 body.  This kind of minimalist perspective has been employed successfully
to study, for example, the circadian rhythms of the avian chick
eye and the dynamics of coupled microwave
oscillators.\cite{Erika,Erika2,Wirkus}

Minimalist models of this sort are not intended to suggest the
mechanisms that underly a given phenomenon but rather to gain a
qualitative understanding of the dynamics of the existing
mechanisms (especially in situations, like this one, in which they
are not known or poorly understood). We also hope that our work
will spur the data collection necessary to develop more realistic
mathematical models that can be tied closer to the application at
hand.  Mathematical studies have the potential to help our
understanding of bipolar disorder considerably, but a lot more
work must be done to reach that stage.  This paper is intended to
be a first step along that path.


\section{Limit-cycle Oscillators for Bipolar II Individuals}
\indent

For modeling purposes, some simplifying assumptions are necessary.
First, although bipolar II disorder can be somewhat erratic, episodes
are known to exhibit recurrent patterns. For a group of patients
with this disorder, there is a periodicity governing the manic and
depressive episodes.\cite{Kupers} Moreover, it is commonly assumed
that the disorder will progress severely if left untreated.

As previously mentioned, we are not attempting to provide an
explanation of the underlying mechanisms of bipolar disorder
(although we hope that we can provide a stepping-stone toward
achieving this highly desirable objective). Rather, our immediate
goal is to better understand the dynamics of the mood swings by
assuming {\it a priori} the existence of an oscillator or
oscillators that might approximate the observed behavior of
bipolar individuals.

Figure 2 in early work by Wehr and Goodwin suggests that the
behavior of the single bipolar patient they studied can be
described qualitatively as a limit cycle oscillator.\cite{Wehr1}
Moreover, a typical bipolar patient undergoes roughly four symptom
changes per year, which corresponds roughly to observations of
Wehr and Goodwin.\cite{Judd,Wehr1,Management1} The rapid cycling
that occurs under the treatment of Desipramine Hydrochloride also
lends itself to a mathematical interpretation in terms of limit
cycle oscillators.  Our analysis provides insight into how these
limit cycle oscillations are induced with treatment.

As hypomanic and depressive episodes are periodic and---if
untreated---increase in severity over time, one can model the mood
and mood change of an untreated individual using a negatively
damped harmonic oscillator,
\begin{equation}\label{Negatively Damped Harmonic Oscillator}
    \ddot{x} - \alpha \dot{x} + \omega^2 x = 0\,,
\end{equation}
where $x$ denotes the emotional state of the patient, $\dot{x}$ is
the rate of mood changes between hypomania and severe depression,
and $\alpha > 0$ and $\omega$ are parameters.  The main drawback
of such a simple model is its unbounded oscillations.

Most patients with bipolar II disorder are diagnosed when they are
in a depressive episode, as hypomania ordinarily does not prevent
the normal function of individuals. (Mania, in contrast, causes 
significant occupational and social discomfort.)  In many cases, hypomania can
even enhance short-term functionality.\cite{Management1}  Because
it often takes several years for bipolar disorder to be properly
diagnosed, we assume treatment begins when a patient is in his
early 20s.  We also suppose that treatment can be modeled using
an autonomous forcing function (although nonautonomous forcing,
involving time-periodic functions such as a sequence of delta
functions or trigonometric functions, is also worth considering).
The inclusion of this forcing term turns equation (\ref{Negatively
Damped Harmonic Oscillator}) into the well-known van der Pol
oscillator,
\begin{equation}\label{vdpol}
    \ddot{x} - \alpha \dot{x} + \omega^2 x = \beta x^2 \dot{x}\,.
\end{equation}

The forcing function $g(x,\dot{x}) = \beta x^2\dot{x}$ represents
aggregate treatment and includes a combination of antidepressants,
mood stabilizers, psychotherapy, and either antipsychotics or
tranquilizers. Finding the correct way to treat a given patient
may take as many as ten to fifteen trials on different
medications. The most comprehensive treatment, however, involves
both medication and psychotherapy.\cite{fawcett}

One can suppose equation (\ref{vdpol}) describes a treated bipolar
individual, as it possesses a unique stable limit cycle
surrounding the origin.\cite{Lienard} It is a specific example of
a more general class of equations of the form
\begin{equation}\label{Lienardform} 
  \ddot{x} + f(x)\dot{x}+h(x)=0\,,
\end{equation}
which is known as the Li\'enard equation. The presence of a
limit cycle indicates that after treatment, the bipolar patient's
mood variations (asymptotically) oscillate within a range of $x$
values to be determined by the parameters in equation
(\ref{vdpol}).

Although this simplistic model gives some insight into the
dynamical properties of periodic mood variations of bipolar
individuals, the characterization of untreated individuals is not
realistic because of the unbounded oscillations (mood swings) that
result from every initial condition. We thus seek a model that not
only captures the behavior of treated bipolar II individuals, but
also does a better job of capturing the dynamics of untreated
individuals.


\subsection{Model 1: van der Pol Oscillator with Autonomous
Forcing} \indent

In this model, we use a van der Pol oscillator\cite{Lienard} to
represent an untreated bipolar II individual.  This
characterization is more realistic than that in equation
(\ref{Negatively Damped Harmonic Oscillator}), as the mood swings
of an untreated patient can still grow large but now approach a
bounded limit cycle rather than eventually becoming infinitely
severe. We again apply treatment in the form of an autonomous
forcing function $g(x, \dot{x})$,
\begin{equation}\label{medicated}
    \ddot{x} - \alpha \dot{x} + \omega^{2}x -
\beta x^{2} \dot{x} = g(x,\dot{x})\,.
\end{equation}
In our analysis, we used
\begin{equation}\label{medgen}
    g(x,\dot{x}) = \gamma x^{4}\dot{x} + \delta x^{2} \dot{x}\,.
\end{equation}
For our simulations, we considered the case $\delta = 0$, but one
can apply a medication of the more general form (\ref{medgen}) to
adjust the coefficient $\beta$.

\begin{figure}
\center
\includegraphics [angle=0, width=.7\textwidth]{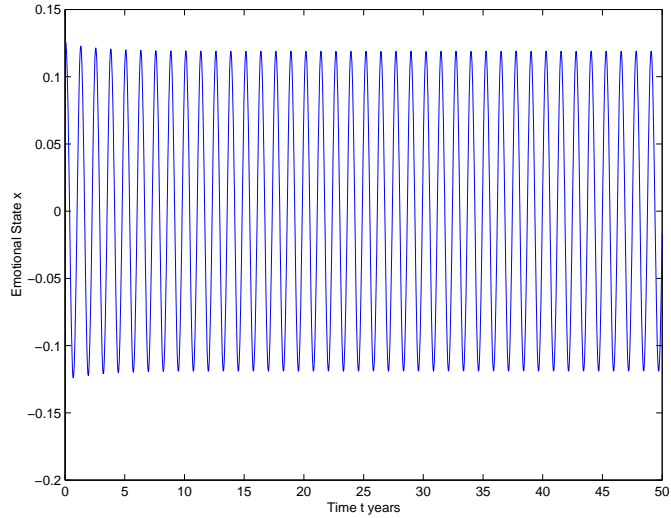}
\caption {Time series representing the behavior of an untreated
patient. The parameters are $\alpha = .36$, $\beta = -100$, and
$\omega = 5$.  This leads to a stable limit cycle with an
amplitude of about $.12$, which we designate as larger than that
describing `normal' mood swings.}\label{graph3}
\end{figure}

\begin{figure}
\center
\includegraphics [angle=0, width=.7\textwidth]{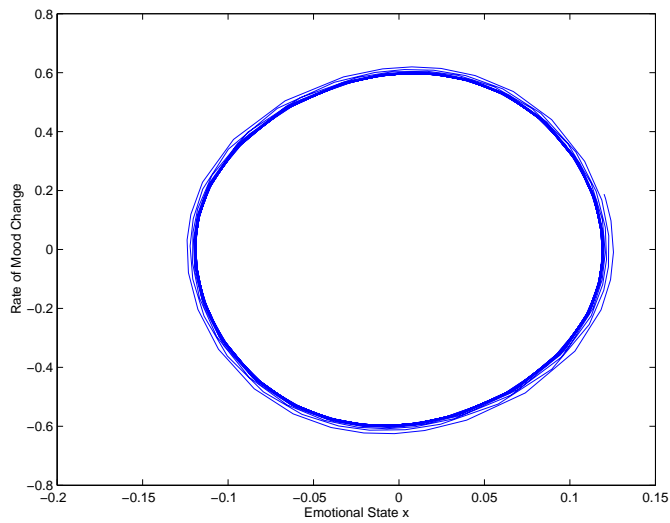}
\caption {Limit cycle corresponding to the time series in
Figure~\ref{graph3}.  This describes the relationship between the
individual's moods and how fast they change.}\label{graph4}
\end{figure}

Because normal individuals also have mood swings, one must
designate how severe such mood variation must be in order for
someone to be diagnosed as bipolar.  In other words, only limit
cycles with some minimal amplitude (say, $0.1$) correspond to
bipolar mood swings. In an untreated patient described by
(\ref{medicated}) with $g(x) \equiv 0$, the parameters $\alpha =
.36$, $\beta = -100$, and $\omega = 5$ lead to a limit cycle with
an amplitude of approximately $0.12$, as shown in Figures \ref{graph3} 
and \ref{graph4}.  In the context of the present model, such an amplitude 
threshold dividing functional individuals from bipolar II ones is arbitrary, 
so the scales in (\ref{medicated}) can be adjusted to account for
individuals who suffer from bipolar disorder to varying degree
(e.g., individuals who have mania rather than hypomania).

\begin{figure}
\center
\includegraphics [angle=0, width=.7\textwidth] {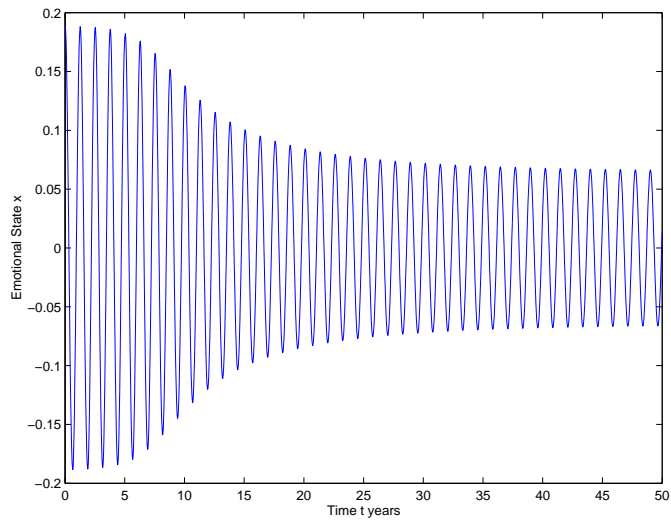}
\caption {Treated patient with parameters $\alpha=.1$,
$\beta=-100$, and $\gamma = 5000$, which yields an unstable limit
cycle with amplitude greater than $.1$ and a stable limit cycle
with amplitude less than $.1.$ This patient begins to receive
treatment at about age $5$ or $10$.  Individuals usually first become
bipolar between the ages of $18$ and $22$, but the onset of bipolar
disorder can occur in early childhood.\cite{fawcett}.}
\label{graph5}
\end{figure}

\begin{figure}
\center
\includegraphics [angle=0, width=.7\textwidth] {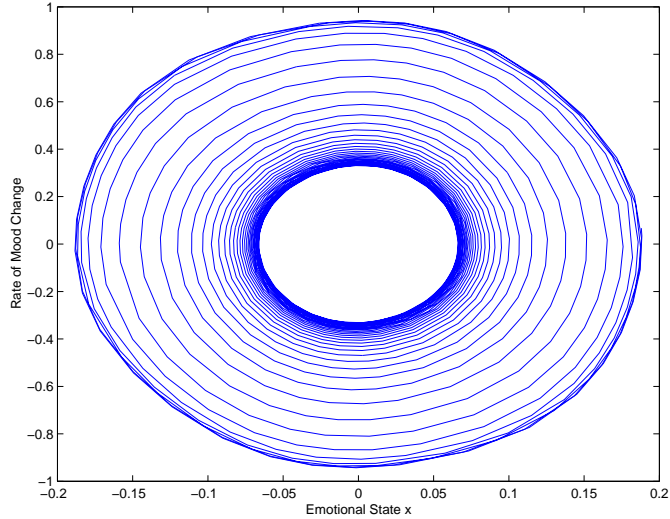}
\caption {Limit cycles corresponding to the time series in
Figure~\ref{graph5}.  The depicted trajectory approaches the
smaller (stable) limit cycle as the time $t$ increases.  The amplitude of
this limit cycle is less than $0.1$, so it describes the mood swings
of a functional individual.}\label{graph6}
\end{figure}

Starting with the van der Pol oscillator caricature of the
untreated patient, we apply the forcing function $g(x,\dot{x})$ to
represent medication. We use (\ref{medicated}) to represent a
treated bipolar II patient, who possesses an unstable, large limit
cycle (greater than the $.1$ amplitude, which is the threshold
that we have chosen for `normal' behavior) and a stable, small
limit cycle. Choosing $\alpha=.1$, $\beta = -100$, and $\gamma=5000$
yields the plots in Figures \ref{graph5} and \ref{graph6}.

Using the method of averaging,\cite{rand} we study the number and
amplitude (denoted $A$) of the limit cycles of equation
(\ref{medicated}).  The dynamics of $A$ are described by {\it slow
flow} equations of motion, whose non-zero equilibria correspond to
limit cycles of equation (\ref{medicated}). Depending on the
values of the parameters, there can be zero, one, or two limit
cycles. The slow flow equations for the amplitude contain both
positive
 and negative equilibria, but only the non-negative solutions are relevant
to the original dynamical system.  Limit-cycle amplitudes must
satisfy
\begin{equation}\label{limitcyc}
    A^{2} = -\frac{\beta}{\gamma} \pm \frac{1}{\gamma}\sqrt{\beta^2 - 8\alpha\gamma}\,.
\end{equation}
Once we find the limit cycle amplitudes, we compute the Jacobian
(in this case, the second derivative) to determine their stability
as functions of $\alpha$, $\beta$, and $\gamma$.  There is always
a slow-flow equilibrium at $A = 0$, which corresponds to the
equilibrium at $(0,0)$ of (\ref{medicated}).  We obtain the
following conditions for the existence and stability of limit
cycles:

\begin{enumerate}

\item Zero limit cycles occur when either
$\beta^2 - 8\alpha\gamma < 0$ (so that $A^2$ is not real) or
$\alpha$, $\beta$, and $\gamma$ all have the same sign (so that
$A^2 < 0$).  In these cases, there are no slow-flow equilibria
with positive $A$.

\item One limit cycle:
\begin{itemize}

\item There is one limit cycle when $\beta/\gamma > 0$ and
$\beta^{2} = 8 \alpha \gamma$. There is a bifurcation at this
point (corresponding to the coalescence of two limit cycles), and
stability cannot be determined by computing a Jacobian.

\item There is one limit cycle when
$\sqrt{\beta^2 - 8\alpha\gamma} > |\beta|$, regardless of the sign
of $\beta/\gamma$. This occurs exactly when $\alpha \gamma < 0$.
Such a limit cycle is always unstable.

\item There is one limit cycle when $\alpha = 0$ and $\beta/\gamma < 0$. In
this situation, the Jacobian is $-\beta^{2}/ 2\gamma$, so the
limit cycle is stable if and only if $\gamma > 0$.
\end{itemize}

\item Two limit cycles occur if $\beta/\gamma < 0$ and
$\beta^2 > 8 \alpha \gamma > 0$. This happens for $\gamma,\,\alpha >
0$ when $\beta < 0$ and for $\gamma, \, \alpha < 0$ when $\beta > 0$.
In the biologically relevant situation describing a treated
individual, the smaller limit cycle is stable.  An example of this
is depicted in Figures \ref{graph5} and \ref{graph6} with the parameter 
values $\alpha = .1$, $\beta = -100$, and $\gamma = 5000$.

\end{enumerate}

\bigskip


\subsection{Model 2: Li\'enard Oscillator with Autonomous Forcing}
\indent

The introduction of a forcing function (representing treatment)
 to Model I results in smaller mood swings; in so doing, however, a larger
unstable limit cycle was introduced.  This has the potential
drawback of predicting that if an individual goes too long without
being diagnosed (and thus the amplitude of the mood swings is too
large), one would need to ensure that the initial condition when
treatment begins is within the basin of attraction of the smaller
stable limit cycle. In other words, if the mood swings are too
large, it might be necessary to drastically reduce the mood
amplitude before introducing `normal' treatment.

Model 2 provides an alternative minimalist framework to study
bipolar II disorder that does not have this drawback. In this
situation, we will demonstrate the presence of three limit cycles.
We will also show that the ones with the largest and smallest
amplitude are stable, whereas the limit cycle between them is
unstable. Toward this end, we consider an equation of the form
\begin{equation}\label{ForcedLienardform}
    \ddot{x} +f(x)\dot{x}+h(x) = g(x,\dot{x})\,,
\end{equation}
where
\begin{equation}
    g(x,\dot{x})=\rho \dot{x}^3+\mu\dot{x}^5+\nu\dot{x}^{11}\,.
\end{equation}
For this model, we consider the constant function $f(x)=-0.38$ and
the linear function $h(x)=180x$ together with the parameters
$\mu=0.78$ and $\nu=-0.00093$. With $\rho=0.38$, we obtain a large
stable limit cycle with mood amplitude approximately equal to
$0.44$. As with Model 1, an untreated patient has bounded mood
swings.

\begin{figure}
\center
\includegraphics [angle=0, width=.7\textwidth] {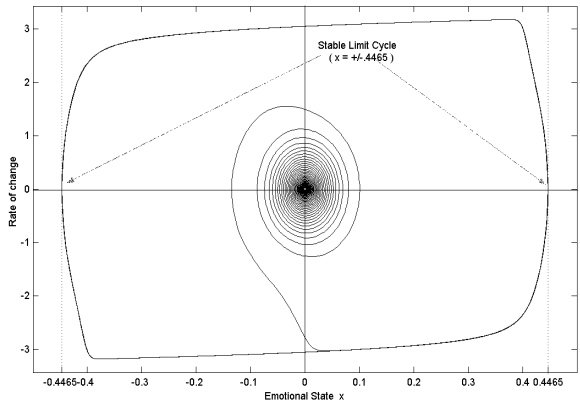}
\caption {Untreated bipolar patient for Model 2.  Only a large
stable limit cycle is present.}\label{PhaseUt}
\end{figure}

\begin{figure}
\center
\includegraphics [angle=0, width=.7\textwidth] {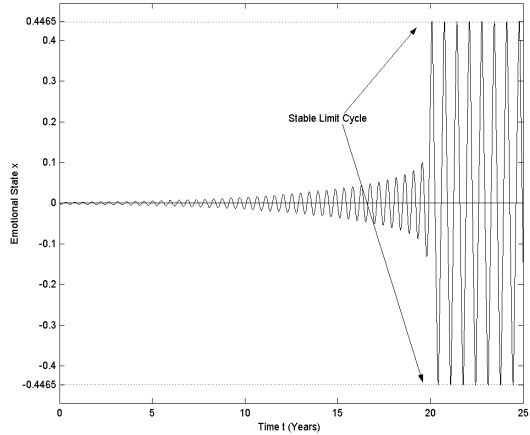}
\caption {Time series corresponding to Figure \ref{PhaseUt}.  The
initial condition lies inside the limit cycle and thus represents
an individual who gradually worsens over time.}\label{TP2}
\end{figure}

Without treatment, this model describes an individual with
steadily worsening mood swings throughout his childhood and
adolescent years. At approximately age $20$, the individual's mood
variations increase dramatically to the point where the individual
can be clinically diagnosed with bipolar II
 disorder.  For the given model and {\it any} initial conditions, the
individual's mood variations settle asymptotically to the stable
limit cycle with mood swing amplitude $|x| \leq 0.45$, as shown in
Figure \ref{PhaseUt}. As this system's limit cycle is globally
stable, every trajectory eventually spirals toward it, yielding
the time series depicted in Figure \ref{TP2}. Observe that the mood varies 
between $\pm 0.4465$.  Moreover, the bipolar II individual described by 
this time series can diagnosed with both hypomania and severe depression 
at age 20.

\begin{figure}
\center
\includegraphics [angle=0, width=.7\textwidth] {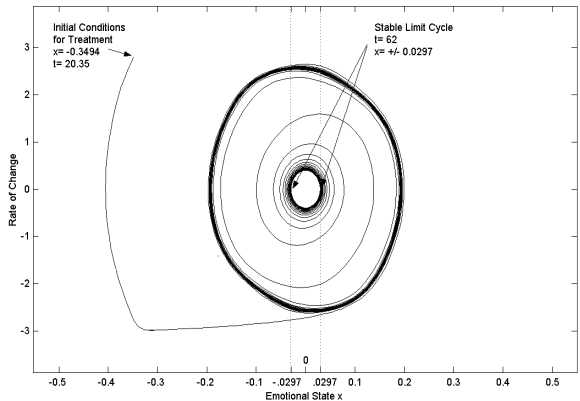}
\caption {Treated bipolar II patient for Model 2.  There is a
large
  outer limit cycle which is stable and an unstable limit cycle lying
  just inside it.  There is also a smaller stable limit cycle inside the
  unstable one; this orbit represents the desired mood swings of a treated
patient.}\label{PhaseT}
\end{figure}

In considering the role of treatment $g(x,\dot{x})$ in this model,
we note that $\rho$ is the key parameter that will be adjusted.
For a treated patient, the value $\rho = -3.302$ (for example)
yields qualitatively appropriate dynamics.  A treated bipolar II
patient is then modeled by
\begin{equation}\label{Bounded Motion}
    \ddot{x} - 0.38 \dot{x} + 180 x = -3.302 \dot{x}^3 + 0.78 \dot{x}^5
- 0.00093 \dot{x}^{11}\,.
\end{equation}
This model encompasses all aspects of treatment as a single
function $g(x, \dot x)$.  In Figure \ref{PhaseT}, we see that by
inputting the new initial conditions into the model that includes
treatment, the patient exhibits favorable emotional patterns.
There is a large stable limit cycle and an unstable limit cycle
just inside it. A smaller stable limit cycle is the desired
emotional pattern of the patient; all initial conditions lying
inside the large unstable cycle yield solutions that spiral toward
this smaller cycle, whose amplitude is sufficiently small so that
the associated mood swings are deemed normal.  One gets the same
qualitative behavior with smaller $|\rho|$ (such as $|\rho| = 2$),
although the separation between the unstable limit cycle and the
larger stable limit cycle surrounding it becomes larger as
$|\rho|$ is decreased.  The chosen value of $\rho$ yields limit
cycles that are almost on top of each other; increasing $|\rho|$
destroys this qualitative structure.

\begin{figure}
\center
\includegraphics [angle=0, width=.7\textwidth] {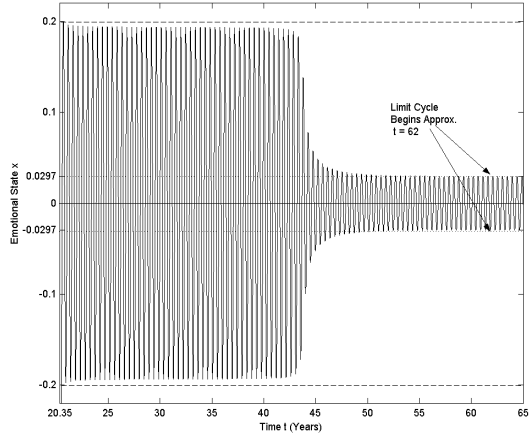}
\caption {Time series corresponding to Figure \ref{PhaseT}.  The
initial condition lies inside the unstable limit cycle and is thus
in the basin of attraction of the smaller stable limit
cycle representing normal mood swings.}\label{TP6}
\end{figure}

The larger limit cycle prevents a treated patient from having
unbounded mood variations that could otherwise occur, for example,
if a perturbation were to make the current mood amplitude too
large. The smaller stable limit cycle represents the desired
low-amplitude mood variations. Comparing Figures \ref{TP2} and
\ref{TP6} reveals that after treatment, the hypomanic and severe
depressive episodes have both disappeared. At the beginning of
treatment, the amplitude of the mood swings stays relatively close
to $0.2$. It then tends asymptotically to the stable limit cycle,
where the mood $x$ varies between approximately $\pm 0.03$. An
important feature of the model demonstrated in Figure \ref{PhaseT}
is that treatment can begin at any time because the small
 limit cycle attracts all solutions that lie inside the unstable limit cycle.
 If the mood swings of the bipolar individual are initially too large, it may
 again be necessary for a brief drastic treatment to bring them into the
basin of attraction of the smaller limit cycle before applying
normal medication.

\section{Bipolar Patients as Interacting Oscillators}
\indent

Examining the possible effects two patients have on each other
while undergoing treatment is of significant mathematical
interest.  Moreover, as mentioned previously, there are known
cases of family units with multiple bipolar II individuals.  Thus,
the interaction of two bipolar patients is also of psychological
interest.\cite{weiss}  It is known to occur, for example, in
households with bipolar sibling pairs.\cite{sibling}  There has
also been some work on using group treatment as a prophylaxis with
patients in remission in order to prevent further
episodes.\cite{colom}  We again stress, however, that clinical
data needs to be collected to examine these ideas more
extensively.


Consider, for example, two treated bipolar II individuals, each of
whom is separately described by equation (\ref{vdpol}).  We transform 
into polar coordinates to provide a natural manner of
adding coupling terms to represent the interaction between the two
patients.  Denoting by $K_j(r_1,\theta_1,r_2,\theta_2)$ the
interactions between the patients, the dynamical system of
interest is
\begin{align}
    \dot{r_1}&= -\beta r_1^3 (\cos{\theta _1})^2(\sin{\theta
_1})^2\nonumber \\ &\quad + r_1 \sin{\theta _1}{\left [\alpha
\sin(\theta_1) + \cos(\theta_1) - \omega^2 \cos(\theta_1) \right
]} +
K_1(r_1,\theta_1,r_2,\theta_2) \,, \notag \\
    \dot{\theta_1} &= \alpha \sin(\theta_1)\cos(\theta_1)-
(\sin{\theta_1})^2\nonumber \\ &\quad -\omega^2(\cos{\theta_1})^2-
\beta r_1^2 (\cos{\theta_1})^3\sin(\theta_1) +
K_2(r_1,\theta_1,r_2,\theta_2) \,,
\notag \\
    \dot{r_2} &= -\beta r_2^3\cos{\theta _2}^2(\sin{\theta _2})^2
\nonumber \\ &\quad + r_2\sin{\theta _2}{\left [\alpha
\sin(\theta_2)+\cos(\theta_2)-\omega^2 \cos(\theta_2) \right ]} +
K_3(r_1,\theta_1,r_2,\theta_2) \,, \notag \\
    \dot{\theta_2} &= \alpha \sin(\theta_2)\cos(\theta_2)-
(\sin{\theta_2})^2\nonumber
\\ &\quad -\omega^2(\cos{\theta_2})^2- \beta r_2^2
(\cos{\theta_2})^3\sin(\theta_2) +
K_4(r_1,\theta_1,r_2,\theta_2)\,. \label{couplepolar}
\end{align}

For simplicity, we assume that the phase difference affects only
the phase terms and the amplitude difference affects only the
amplitude terms.  We also assume linear proportionality, so that
the coupling terms simplify to
\begin{align}
    K_1 &= k_1(r_2-r_1)\,, \quad K_2 = k_2(\theta_2-\theta_1)\,,
\notag \\
    K_3 &= k_3(r_1-r_2)\,, \quad K_4 = k_4(\theta_1-\theta_2)\,,
\label{coupling1}
\end{align}
where the $k_i$ are the proportionality constants.  An important
detail to note is that the angular coupling terms in
(\ref{coupling1}) include no limitation on the phase angle, so
that $\theta_1 - \theta_2$ can increase beyond $2\pi$. There is no
difficulty, however, when we convert back to rectangular
coordinates.  The coupling terms remain bounded, as the radial
coupling terms are proportional to $\sqrt{x_i^2 + \dot{x}_i^2}$
and the angular ones are proportional to $\arctan(\dot{x}/x)$.
When we numerically integrate these dynamical equations, we
utilize $\arctan\,(\dot{x}/x) \in (-\pi/2, \pi/2)$. Although the
arctangents in rectangular coordinates create a discontinuous
vector field, the behavior near the discontinuities correctly
models the biology observed in situations in which two patients
with bipolar disorder are interacting.  (That is, we observe
fluctuations near in-phase and out-of-phase modes).

Other choices of coupling terms avoid unbounded phase differences.
Consider, for example,
\begin{align}
    K_1 &= k_1(r_2-r_1)\,, \quad K_2 = k_2\sin(\theta_2-\theta_1)\,,
\notag \\
    K_3 &= k_3(r_1-r_2)\,, \quad K_4 = k_4\sin(\theta_1-\theta_2)\,,
\label{coupling2}
\end{align}
where the $k_i$ are again proportionality constants. We have
conducted some mathematical analysis of (\ref{couplepolar}) with both 
(\ref{coupling1}) and (\ref{coupling2}) and discuss one of our results 
briefly. Note, however, that although these results are mathematically
interesting, we only mention these models in passing because
insufficient clinical data is available to properly constrain the
numerous forms of coupling available when constructing mathematical models
describing interacting bipolar II individuals.  Consequently, we
only state results for two common types of solutions in such
systems.

In coupled systems, important types of behavior include in-phase
modes, for which $x_1=x_2$ and $y_1 = y_2$, and out-of-phase
modes, for which $x_1 = -x_2$ and $y_1 = -y_2$.  For the first
form of coupling (\ref{coupling1}), both in-phase and out-of-phase
modes exist and are stable with relatively small basins of
attractions. Larger stable motions exist that
 attract initial conditions starting outside those basins of attraction.  Both
in-phase and out-of-phase modes also exist with the second form of
coupling (\ref{coupling2}), and the former appears to be the only
stable motion in the
 system.  Initial conditions starting far away from the in-phase mode
eventually approach this mode.

Interpreted biologically, two coupled bipolar II individuals with
similar moods ($x_1 \approx x_2$) and mood cycling rates ($y_1
\approx y_2$) tend to remain in-phase. They are synchronized in
the sense that they enter hypomanic and depressive states
(roughly) concurrently. The same argument applies to out-of-phase
modes, so if two bipolar II individuals are almost completely
out-of-phase, they will remain that way. One individual enters a
hypomanic phase when the other enters a depressive phase.


\section{Discussion}
\indent

In this work, we proposed two limit cycle oscillator models that
attempt to explain qualitatively the moods swings of bipolar II individuals. 
 There is insufficient data at this time to construct a quantitatively
detailed model describing bipolar disorder, so this first step
employing minimalist models is an essential one.  Indeed, similar
phenomenological models have yielded significant insight into the
dynamics of several oscillatory biological and physical
phenomena.\cite{Erika,Erika2,Wirkus}

Even without time-series data to describe the moods of bipolar patients,
the models we have developed could prove important in increasing
the understanding of the effect of treatment on the cyclic
behavior of bipolar individuals. Because the two-dimensional
dynamical systems we utilize provide simple caricatures of the
behavior of bipolar individuals, it is concomitantly easy to
analyze these models. Moreover, despite their simplicity, they
successfully capture qualitatively the known behavior of treated
and untreated bipolar II individuals.  A quantitative analysis of
their dynamics, which is our long-term goal, will require refining
our models using time-series data from clinical trials.  It is our
hope that the present work will stimulate further mathematical
treatments that incorporate such data as well as the data
collection that will permit these modeling efforts.

In Model 1, we considered treatment as a forcing function that led
to a stable limit cycle with smaller amplitude, which we
interpreted as a decrease in the severity of the mood swings of
the bipolar patient.  This model suggests that individuals that
are not diagnosed at a sufficiently young age might need drastic
intervention to bring their mood swings to a reasonable level.
Given that individuals are usually diagnosed in the depressive
state and that bipolar disorder is frequently misdiagnosed initially, 
this aspect of Model 1 is especially noteworthy.\cite{vornik,iain}

Model 2 provides an alternative framework in which a bipolar II
individual possesses two stable limit cycles even without
treatment; satisfactory treatment then confines the patient to the
one with smaller amplitude.  The rapid cycling induced by Desiramine 
Hydrochloride that was reported by Wehr and Goodwin\cite{Wehr1} can be 
interpreted as a slight increase in the amplitude of the
 oscillation but---more centrally---as an increase in the oscillation
frequency.  This phenomenon can be addressed within
the modeling framework we have developed by allowing the frequency
parameter to also be perturbed slightly by treatment.

At this point, it is also important to briefly consider how our
models can be both refined and generalized.  First,
incorporating biological data is of paramount importance. We used
minimal data because little data is available in a form that
facilitates the creation or refinement of mathematical models.
This is the largest deficiency in our analysis; employing our
modeling framework along with data from psychological testing of
bipolar II individuals will yield a significantly increased
understanding of this medical condition.  In fact, one of the primary 
purposes of this work is to motivate the collection of such data so that
detailed models of bipolar disorder can ultimately be developed.
In addition to time-series data describing the behavior of a large
number of bipolar patients, it is also desirable to have data
describing the mood variations of ``normal'' individuals to
provide a basis for comparison.

Numerous generalizations of our models can be considered to
further examine mood disorders using limit-cycle oscillators. For
example, one can incorporate the fact that the medication administered to 
bipolar II individuals typically take several days to reach therapeutic 
levels in the blood stream by adding a time-delay to the treatment function. 
One can also gain further mathematical and biological insight by directly 
examining the bifurcations that occur in our models. Moreover, one can
incorporate explicit time-dependence (for example, via delta
functions representing instantaneous medication) into the
treatment function by considering a nonautonomous forcing function
$g(x,t)$.

Limit cycle oscillators can also be used to study behavioral
cycling in mood disorders besides bipolar disorder, as such phenomena are of
considerable interest to both psychologists and
psychiatrists.\cite{iain} Pertinent disorders include unipolar
depression and cyclothymia. Individuals suffering from the latter
disorder rapidly flip back and forth between depressed and
euphoric moods. Their high and low periods, which are relatively
short, are much less intense than those of bipolar individuals.
Although it is not known precisely what leads individuals from one
pole to the other, possible causes include psychosocial stress,
disrupted sleep patterns, or some endogenous or internal
biological mechanism that is not connected to environmental
stimuli.\cite{riso} It is also not known what happens to cycling
over time.  It is speculated that an individual's cycling may slow
down with age. From a mathematical perspective, we wish to
highlight that the effects of disrupted sleep patterns can be
studied by coupling a cyclothymic (or bipolar) individual with a
limit-cycle oscillator representing the human sleep-wake
cycle.\cite{sleep}

Another important issue to consider is the threshold for flipping
poles. It has been theorized, for example, that the threshold for
depression (and for flipping poles) gets lower over time.  It has
been speculated (by Robert Post\cite{post2}, for example) that the
longer one is ill, the more autonomous or disconnected episodes
become from the environment. This is termed ``kindling theory''
and is believed to be analogous to what happens in epilepsy, in
which there is a gradual kindling of biological disturbances in
the brain that eventually surpasses the threshold for a seizure.


\section{Conclusions}

In this paper, we provide a mathematical framework for the
modeling of bipolar disorder in terms of low-dimensional limit
cycle oscillators. We propose and analyze two phenomenological
models of bipolar individuals. Rather than focusing on the method
of diagnosis, which is a difficult medical problem, we instead
concentrate on the dynamics of our models under the proposed
treatment strategy, which involves a combination of drugs and
therapy.  The purpose of our modeling efforts is not to suggest a specific
treatment for bipolar individuals but rather to provide some
insight into the complicated dynamics of bipolar disorder.  In
this respect, we view our work as a first step in developing
mathematical models of the mood swings of bipolar individuals.
Our intent is to provide a mathematical framework that ultimately
leads to the development of more detailed models of bipolar
disorder
 that incorporate clinical data.  With this work, we hope to motivate the
collection of time-series data from clinical trials that will lead
to refinements of our model that incorporate such data.  In our
view, dynamical systems theory and mathematical modeling in
general can lead to important advancements in the understanding of
bipolar disorder.


\section*{Acknowledgments}
\indent

Valuable discussions with Michael Stubna, Carlos Castillo-Chavez,
Richard Rand, Abdul-Aziz Yakubu, John Franke, Roxana Lopez-Cruz,
Terry Kupers, Dane Quinn, Richard Frenette, Larry Riso, and Iain
Macmillan are gratefully acknowledged. This research has been
partially supported by grants given by the National Science
Foundation, National Security Agency, the Sloan Foundation
(through the Cornell-Sloan National Pipeline Program in the
Mathematical Sciences), and the NSF VIGRE program at Georgia Tech.
Substantial financial and moral support was also provided by the
Office of the Provost of Cornell University, the College of
Agriculture \& Life Science
 (CALS), and the Department of Biological Statistics \& Computational Biology.



\begin{thebibliography}{1}

\bibitem{Diagnostic}
American Psychiatric Association (2000), Diagnostic and
statistical manual of mental disorders : DSM-IV-TR. American
Psychiatric Association, Washington DC.

\bibitem{sibling}
P. Bennett, R. Segurado, {\it et al.} (2002),
\newblock{The Wellcome trust UK-Irish bipolar affective disorder sibling-pair
genome screen: first stage report.}
\newblock{\it Molecular Psychiatry}, 7: 2, 189-200.

\bibitem{Erika}
Erika T. Camacho (2003),
\newblock{Mathematical Models of Retinal Dynamics}
\newblock{\it Ph.D. Thesis}, Center for Applied Mathematics.
\newblock Cornell University, Ithaca, NY.

\bibitem{Erika2}
Erika T. Camacho, Richard H. Rand, and Howard H. Howland (2004)
\newblock{Dynamics of Two van der Pol Oscillators Coupled via a Bath}
\newblock{\it Special Boley Issue of the International Journal of Solids and
Structures}, 2004.

\bibitem{colom}
Francesco Colom, Eduard Vieta, {\it el al.} (2003),
\newblock{A randomized trial on the efficacy of group psychoeducation in the
prophylaxis of recurrences in bipolar patients whose disease is in
remission.}
\newblock{\it Archives of General Psychiatry}, 60: 4, 402-407.

\bibitem{ehlers}
{Cindy L. Ehlers (1995)},
\newblock{Chaos and complexity: Can it help us to understand mood and
behavior?}
\newblock{\it Arch Gen Psych}, 52: 960-964.

\bibitem{fawcett}
Jan Fawcett, M.D., Bernard Golden, Ph.D., and Nancy Rosenfeld
(2000),
\newblock {\it New Hope for People with Bipolar Disorder},
\newblock Prima Publishing, Roseville, CA.

\bibitem{iain}
{I. Nicol Ferrier, Iain C. Macmillan, and Allan H. Young (2001)},
\newblock{The search for the wandering thymostat: a review of some
developments in bipolar disorder research},
\newblock {\it British Journal of Psychiatry}, 178: 41, s103-s106.

\bibitem{bip}
Paula Ford-Martin (1995),
\newblock {\it Bipolar Disorder}.
\newblock Gale Encyclopedia of Alternative Medicine, January 01.

\bibitem{Management2}
Ellen Frank, Holly A. Swartz, and David J. Kupfer (2000),
\newblock Interpersonal and Social Rhythm Therapy: Managing the Chaos
of Bipolar Disorder.
\newblock{\it Biological Psychiatry}, 48: 6, 593-604.

\bibitem{LimitCylces}
Hector Giacomini and Sebastian Neukirch (1997),
\newblock On the Number of Limit Cycles of the Li\'enard Equation.
\newblock{\it Physical Review E},  56: 4.

\bibitem{DiffEqn}
Stephen W. Good (2000),
\newblock{\it Differential Equations and Linear Algebra, second
edition.}
\newblock{Prentice-Hall Inc. Upper Saddle River, New Jersey, Second
Edition}.

\bibitem{gott}
A. Gottschalk, M. S. Bauer, and P.C. Whybrow (1995),
\newblock{Evidence of chaotic mood variation in bipolar disorder,}
\newblock{{\it Arch Gen Psychiatry}, 52: 947-959.}

\bibitem{Management1}
Kim Griswold (2000),
\newblock {\it Management of Bipolar Disorder}.
\newblock American Family Physician, September 15.

\bibitem{Dstool}
J. Guckenheimer, M. Meyers, F. Wicklin, and P. Worfolk (1991),
\newblock {\it DSTool: Dynamical System Toolkit with an Interactive Graphical
Interface}. \newblock Department of Applied Mathematics, Cornell
University, Ithaca, NY.

\bibitem{treat}
David Haslam, MSc, MD, Sidney Kennedy, MD, FRCPC, Vivek Kusumakar,
MD, FRCPC, Stan Kutcher, MD, FRCPC, Raymond Matte, MD, FRCPC,
Sagar Parikh, MD, FRCPC, Peter Silverstone, MD, FRCPC, Verinder
Sharma, MBBS, FRCPC, Lakshmi Yatham, MD, FRCPC,
\newblock {\it A Summary of Clinical Issues and Treatment Options}.
\newblock Bipolar Disorder Sub-Committee Canadian Network for Mood and
Anxiety Treatments (CANMAT). {\it
http://www.canmat.org/gpsfps/nine/bipolar.html}.

\bibitem{Judd}
Lewis L. Judd, Hagop S. Akiskal, et al. (2003),
\newblock{ A Prospective Investigation of the Natural History
of the Long-term Weekly Symptomatic Status of Bipolar II Disorder}
\newblock{\it Archives of General Psychiatry}, 60: 3, 261-269.

\bibitem{Kupers}
Terry Kupers (June 29, 2002),
\newblock Private Communication,
\newblock Wright Institute.

\bibitem{vornik}
L. Lewis and L.A. Vornik (2003),
\newblock{Perceptions and impact of bipolar disorder: How far have we really
come? Results of the National Depressive and Manic Depressive
Association 2000 survey of individuals with bipolar disorder},
\newblock {\it Journal of Clinical Psychiatry}, 64: 2, 161-174.

\bibitem{iain2}
{Iain C. Macmillan (May 25, 2004)},
\newblock Private Communication,
\newblock Consultant Psychiatrist, Early Intervention Service,
\newblock Norfolk Mental Health Care NHS Trust.

\bibitem{post2}
{Robert M. Post (1992)},
\newblock{Transduction of psychosocial stress into the neurobiology of
recurrent affective disorder},
\newblock {\it American Journal of Psychiatry}, 149, 999-1010.

\bibitem{post}
{Robert M. Post and David A. Luckenbaugh (2003)},
\newblock {Unique design issues in clinical trials of patients with
bipolar affective disorder},
\newblock {\it Journal of Psychiatric Research}, 37: 1, 61-73.

\bibitem{rand}
Richard H. Rand (1994),
\newblock {\it Topics in Nonlinear Dynamics with Computer Algebra},
\newblock Computation in Education: Mathematics, Science and
Engineering Series, vol.~1,
\newblock Gordon and Breach Science Publishers, USA.

\bibitem{riso}
{Lawrence Riso (July 9, 2003)},
\newblock Private Communication,
\newblock Department of Psychology, Georgia State University.

\bibitem{Lienard}
Steven H. Strogatz (1994),
\newblock{\it Nonlinear Dynamics and Chaos: With Applications in
Physics, Biology, Chemistry, and Engineering.}
\newblock{Reading, Mass: Addison-Wesley Publishing Company.}

\bibitem{sleep}
Steven H. Strogatz (1986),
\newblock{\it The Mathematical Structure of the Human Sleep-Wake Cycle}.
\newblock{Lecture Notes in Biomath. {\bf 69}. New York, NY: Springer-Verlag.}

\bibitem{Totterdell}
{Peter Totterdell, Rob B. Briner, Brian Parkinson, and
Shirley Reynolds (1996)}
\newblock{Fingerprinting time series: Dynamic patterns in
self-report and performance measures uncovered by a graphical
non-linear method}.
\newblock{\it British Journal of Psychology}, 87: 43-60.

\bibitem{Wehr1}
{Thomas A. Wehr and Frederick K. Goodwin (1979)},
\newblock{Rapid Cycling in Manic-Depressives Induced by Tricyclic
Antidepressants}.
\newblock{\it Arch Gen Psychiatry}, 36: 555-559.

\bibitem{weiss}
Roger D. Weiss, Margaret L. Griffin, {\it et al.} (2000),
\newblock{Group therapy for patients with bipolar disorder and substance
dependence: Results of a pilot study.}
\newblock{\it Journal of Clinical Psychiatry}, 61: 5, 361-367.

\bibitem{Wirkus}
Stephen A. Wirkus and Richard H. Rand (2002),
\newblock{The Dynamics of Two Coupled van der Pol Oscillators with Delay
Coupling},
\newblock{\it Nonlinear Dynamics}, 30: 3, 205-221.


\end{thebibliography}
\end{document}